\documentclass[12pt]{article}
\usepackage{epsfig,here}
\usepackage{graphicx}
\usepackage{amsmath,amssymb}
\usepackage{latexsym}
\usepackage{bbm}

\def\bdef#1{#1}
\def\bdef#1{{\bf #1}} 
\def\bigvev#1{{\mathbb E}[#1]}

\def\Vol{\mbox{Vol}}

\def\BC{\mathbb{C}}

\def\BZ{\mathbb{Z}}
\def\BR{\mathbb{R}}
\def\BP{\mathbb{P}}

\def\CM {{\cal M}}
\def\CN {{\cal N}}
\def\CR {{\cal R}}

\def\CF {{\cal F}}

\def\CP {{\cal P }}
\def\CL {{\cal L}}

\def\CO {{\cal O}}

\def\CH {{\cal H}}

\def\half{\frac{1}{2}}

\newcommand{\eq}[1]{Eq.~(\ref{eq:#1})}

\renewcommand{\Im}{{\rm Im~ }}
\renewcommand{\Re}{{\rm Re~ }}

\def\one{{\hbox{ 1\kern-.8mm l}}}

\def\zb{\bar{z}}
\def\sb{\bar{s}}
\def\ib{{\bar{i}}}
\def\jb{{\bar{j}}}
\def\kb{{\bar{k}}}
\def\pb{\bar{\partial}}
\def\Dbar{\bar{D}}
\begin{document}

\pagestyle{plain}

\parindent 0mm
\parskip 6pt

\vspace*{-1.8cm}

\begin{center}
{\Large\bf 
Random algebraic geometry, attractors and flux vacua}\\[2mm] 
Michael R. Douglas\\[2mm]
\normalsize NHETC, Rutgers University, Piscataway, NJ 08855 USA\\
{\it and}
\normalsize IHES, Bures-sur-Yvette FRANCE 91440\\
E-mail: mrd@physics.rutgers.edu
\end{center} 

\vspace*{1cm} 

\section{Introduction}

A classic question in probability theory, studied by M. Kac, S. O.
Rice and many others, is to find the expected number and distribution
of zeroes or critical points of a random polynomial.  The same
questions can be asked for random holomorphic functions or sections of
bundles, and are the subject of ``random algebraic geometry.''

While this theory has many physical applications, in this article we
focus on a variation on a standard question in the theory of
disordered systems.  This is to find the expected distribution of
minima of a potential function randomly chosen from an ensemble, which
might be chosen to model a crystal with impurities, a spin glass, or
another disordered system.  Now whereas standard potentials are real-valued
functions, analogous functions in supersymmetric theories, such as
the superpotential and the central charge. are holomorphic sections
of a line bundle. Thus one is interested in finding the distribution
of critical points of a randomly chosen holomorphic section.

Two related and much-studied problems of this type
are the problem of finding attractor points
in the sense of Ferrara, Kallosh and Strominger,
and the problem of finding flux vacua as
posed by Giddings, Kachru 
and Polchinski.  These problems involve a good
deal of fascinating mathematics and are good illustrations of the
general theory.

\section{Elementary random algebraic geometry}

Let us introduce this subject
with the problem of finding the expected distribution
of zeroes of a random polynomial,
$$ 
f(z) = c_0 + c_1 z + \ldots c_N z^N .
$$

We define a \bdef{random polynomial} to be a probability measure on a space
of polynomials.  A natural choice might be independent
Gaussian measures on the coefficients,
\begin{equation} \label{eq:gaussian}
d\mu[f] = d\mu[c_0,\cdots,c_N] = 
\prod_{i=0}^N d^2c_i\ \frac{\sigma_i}{2\pi}\ e^{- |c_i|^2/2\sigma_i^2} .
\end{equation}
We still need to choose the variances.  At first the most natural choice
would seem to be equal variance for each coefficient, say $\sigma_i=1/2$.
We can characterize this ensemble by its two-point function,
\begin{eqnarray*}
G(z_1,\zb_2) 
&\equiv \bigvev{f(z_1) f^*(\zb_2)} = \int d\mu[f]\ f(z_1) f^*(\bar z_2) \\
&= \sum_{n=0}^N (z_1 \zb_2)^n  \\
&=  \frac{1-z_1^{N+1}\zb_2^{N+1}}{1-z_1\zb_2} \label{eq:constvar}
\end{eqnarray*}

We now define $d\mu_0(z)$ to be
a measure with unit weight at each solution of $f(z)=0$, such that
its integral over a region in $\BC$
counts the expected number of zeroes in that region.
It can be written in terms of the standard Dirac delta function, 
by multiplication by a Jacobian factor, 
\begin{equation} \label{eq:expzer}
d\mu_0(z) = \bigvev{\delta^{(2)}(f(z))\ \partial f(z)\ \pb f^*(\zb)} .
\end{equation}
To compute this expectation value,
we introduce a constrained two-point function,
$$
G_{f(z)=0}(z_1,\zb_2) = 
\frac{\bigvev{\delta^{(2)}(f(z))\ f(z_1)\ f^*(\zb_2)}}%
 {\bigvev{\delta^{(2)}(f(z))}}
$$
It could be explicitly computed by using the constraint $f(z)=0$
to solve for a coefficient $c_i$ in the Gaussian integral,
{\it i.e.} projecting on the linear subspace $0 = \sum c_i z^i$.
The result, in terms of $G(z_1,\zb_2)$, is
$$
\bigvev{\delta^{(2)}(f(z))}  = \frac{1}{\pi G(z,\zb)} ; 
$$
$$
G_{f(z)=0}(z_1,\zb_2) = 
 G(z_1,\zb_2) - \frac{G(z_1,\zb)G(z,\zb_2)}{ G(z,\zb)} .
$$
as can be verified by considering
$$
\bigvev{\delta^{(2)}(f(z))\ f(z)\ f^*(\zb_2)} \propto
G_{f(z)=0}(z,\zb_2) = 
 G(z,\zb_2) - \frac{G(z,\zb)G(z,\zb_2)}{G(z,\zb)} = 0
$$
\eq{expzer} follows from this simply by taking derivatives:
\begin{eqnarray*}
d\mu_0(z) &= \frac{1}{G(z,\zb)}
\lim_{z_1,z_2\rightarrow z} D_1 \Dbar_2 G_z(z_1,\zb_2) \\
&= \frac{1}{\pi} \partial \pb \log G(z,\zb) .
\end{eqnarray*}

For the constant variance ensemble \eq{constvar},
\begin{equation} \label{eq:concirc}
d\mu_0(z) = \frac{d^2z}{\pi} 
\left(
 \frac{1}{(1-z\zb)^2} - \frac{(N+1)^2 (z\zb)^N }{ (1-(z\zb)^{N+1})^2} \right) .
\end{equation}
We see that as $N\rightarrow\infty$, the zeroes
concentrate on the unit circle $|z|=1$ (Hammersley, 1954).

A similar formula can be derived for the distribution of roots 
of a real polynomial on
the real axis, using $d\mu(t)=\bigvev{\delta(f(t)) |df/dt|}$.
One obtains (Kac, 1943):
$$
d\mu_0^r(t) = \frac{dt}{\pi} 
\sqrt{
 \frac{1}{  (1-t^2)^2} - \frac{(N+1)^2 t^{2N}}{ (1-t^{2N+2})^2} } .
$$
Integrating, one finds the expected number of real zeroes
of a degree $N$ random real polynomial is $E_N \sim \frac{2}{\pi} \log N$,
and as $N\rightarrow\infty$ the zeroes are concentrated at $t=\pm 1$.

While concentration of measure is a fairly generic property for random
polynomials, it is by no means universal.
Let us consider another Gaussian ensemble, with variance
$\sigma_n = {N!/n!(N-n)!}$.  This choice leads to
a particularly simple two-point function,
\begin{equation} \label{eq:invG}
G(z,\zb) = (1+z\zb)^N ,
\end{equation}
and the distribution of zeroes
\begin{equation} \label{eq:distFS}
d\mu_0 = \frac{1}{\pi}\partial\pb \log G = \frac{N d^2z}{\pi(1+z\zb)^2} .
\end{equation}
Rather than concentrate the zeroes, in this ensemble zeroes are
uniformly distributed according to the volume
of the Fubini-Study ($SU(2)$-invariant)
K\"ahler metric 
$$
\omega = \partial\pb K ; \qquad K = \log (1+z\zb)
$$
on complex projective space $\BC\BP^1$.

We can better understand the different behaviors in our two examples
by focusing on 
a hermitian inner product $(f,g)$ on function space, associated to
the measure \eq{gaussian} by the formal expression
$$
d\mu[f] = [Df] e^{-(f,f)} .
$$
In making this precise, let us generalize a bit further and allow $f$ 
to be a holomorphic section of a line bundle $\CL$, say $\CO(N)$ over
$\BC\BP^1$ in our examples.  We then choose an orthonormal basis
of sections $(s_i,s_j)=\delta_{ij}$, and write
\begin{equation} \label{eq:fsum}
f \equiv \sum_i c_i s_i
\end{equation}
and
$$
d\mu[f] =
\frac{1}{(2\pi)^N} \prod_{i=1}^N d^2c_i\ e^{- |c_i|^2/2} .
$$
We can then compute the two-point function
\begin{equation} \label{eq:twopoint}
G(z_1,\zb_2) 
\equiv \bigvev{s(z_1) s^*(\zb_2)} = 
\sum_{i=1}^N s_i(z_1) s_i^*(\zb_2) .
\end{equation}
and proceed as before.  

In these terms, the simplest way to describe the measure for our first example
is that it follows from the inner product on the unit circle,
$$
(f,g) = \oint_{|z|=1} \frac{dz}{2\pi z} f^*(z) g(z) .
$$
Thus we might suspect that this has something to do with
the concentration of \eq{concirc} on the unit circle.  Indeed, this
idea is made precise and generalized in
(Shiffman and Zelditch, 2003).

Our second example belongs to a class of problems in which $\CM$ is
compact and $\CL$ positive.  In this case, the space $H^0(\CM,\CL)$
of holomorphic sections is finite dimensional, so we can take the
basis to consist of all sections.  Then, if $\CM$ is in addition
K\"ahler, we can derive all the other data from a choice of hermitian
metric $h(f,g)$ on $\CL$.  In particular, this determines a
K\"ahler form $\omega$ as the curvature of the metric compatible connection,
and thus a volume form $\Vol_\omega = \omega^n/n!$.
We then define the inner product to be 
$$
(f,g) = \int_\CM \Vol_\omega\ h(f,g).
$$
Thus, the
measure \eq{gaussian} and the final distribution \eq{expzer} are
entirely determined by $h$.
In these terms, the underlying reason for the simplicity of
\eq{distFS} is that we started with the $SU(2)$ invariant metric $h$, so
the final distribution must be invariant as well.
More generally, \eq{twopoint} is a Szeg\"o kernel.  Taking
$\CL=\CL_1^{\otimes N}$ for $N$ large, this has a known asymptotic
expansion, enabling a rather complete treatment (Zelditch, 2001).

Our two examples also make the larger point that a wide variety of
distributions are possible.  Thus we must put in some information
about the ensemble of random polynomials or sections which appear in
the problem at hand, to get convincing results.

The basic computation we just discussed can be vastly generalized: to
multiple variables, multipoint correlation functions, many different
ensembles, and different counting problems.  We will discuss the
distribution of critical points of holomorphic sections below.

\section{The attractor problem}

We now turn to our physical problems.
Both are posed in the context of compactification of
the type IIb superstring theory on a Calabi-Yau threefold $M$.
This leads to a four dimensional effective field theory with
$N=2$ supersymmetry, determined by the geometry of $M$.

Let us begin by stating the attractor problem mathematically,
and afterwards give its physical background.  We begin by reviewing
a bit of the theory of Calabi-Yau manifolds.
By Yau's
proof of the Calabi conjecture, the moduli space of Ricci-flat metrics
on $M$ is determined by a choice of complex structure on $M$, 
denote this $J$, and a
choice of K\"ahler class.  Using deformation theory, it can be shown
that the moduli space of complex structures, denote this $\CM_c(M)$,
is locally a complex manifold of dimension $h^{2,1}(M)$.  A point $J$ in
$\CM_c(M)$ picks out a holomorphic three-form $\Omega_J\in
H^{3,0}(M,\BC)$, unique up to an overall choice of normalization.
The converse is also true; this can be made precise by defining
the \bdef{period map} $\CM_c(M)\rightarrow \BP(H^3(M,\BZ)\otimes\BC)$
to be the class of $\Omega$ in $H^3(M,\BZ)\otimes\BC$
up to projective equivalence.
One can prove that the period map is
injective (the Torelli theorem), locally in general and globally in
certain cases such as the quintic in $\BC\CP^4$.

Now, the data for the attractor problem is 
a \bdef{charge}, a class $\gamma\in H^3(M,\BZ)$.  An
\bdef{attractor point} for $\gamma$
is then a complex  structure $J$ on $M$ such that
\begin{equation} \label{eq:att}
\gamma \in H_J^{3,0}(M,\BC) \oplus H_J^{0,3}(M,\BC) .
\end{equation}
This amounts to $h^{2,1}$ complex conditions on the $h^{2,1}$ complex
structure moduli, so picks out isolated points in $\CM_c(M)$, the 
attractor points.

There are many mathematical and physical questions one can ask about
attractor points, and it would be very interesting to have a general
method to find them.  As emphasized by G. Moore, this is one of the
simplest problems arising from string theory in which integrality
(here due to charge quantization) plays a central role, and thus it
provides a natural point of contact between string theory and number
theory.  For example, one might suspect that attractor Calabi-Yau's
are arithmetic, {\it i.e.}  are projective varieties whose defining
equations live in an algebraic number field.  This can be shown to
always be true for $K3\times T^2$, and there are conjectures about
when this is true more generally (Moore, 2004).

A simpler problem is to characterize the distribution of attractor points in
$\CM_c(M)$.  As these are infinite in number, one must introduce some
control parameter.  While the first idea which might come to mind is
to bound the magnitude of $\gamma$,
since the intersection form on $H^3(M,\BZ)$ is antisymmetric,
there is no natural way to do this.
A better way to get a finite set is to bound the period of $\gamma$,
and consider the attractor points satisfying
\begin{equation} \label{eq:bound}
Z_{max}^2 \ge |Z(\gamma;z)|^2 \equiv
\frac{|\int_M \gamma\wedge \Omega|^2}{\int_M \Omega\wedge\bar\Omega}.
\end{equation}
As an example of the type of result we will discuss below,
one can show that for large $Z_{max}$, the 
density of such attractor points asymptotically approaches the Weil-Peterson
volume form on $\CM_c$.

We now briefly review the origins of this problem, in the physics of
$1/2$ BPS black holes in $N=2$ supergravity.  We begin by introducing
local complex coordinates $z^i$ on $\CM_c(M)$.  Physically, these can
be thought of as massless complex scalar fields.  These sit in vector
multiplets of $N=2$ supersymmetry, so there must be $h^{2,1}(M)$
vector potentials to serve as their bosonic partners under
supersymmetry.  These appear because the massless modes of the type IIb
string include various higher rank $p$-form gauge potentials, in
particular a self-dual four-form which we denote $C$.  
Self-duality means that
$dC=*dC$ up to non-linear terms, where $*$ is the Hodge star
operator in ten dimensions.

Now, Kaluza-Klein reduction of this four-form potential produces
$b^3(M)$ one-form vector potentials $A_I$ in four dimensions.  Given an
explicit basis of three-forms $\omega_I$ for 
$H^3(M,\BR)\cap H^3(M,\BZ)$, this
follows from the decomposition
$$
C = \sum_{I=1}^{b_3} A_I \wedge \omega_I + {\rm massive~ modes}.
$$
However, because of the self-duality relation, only half of
these vector potentials are independent; the other half are determined
in terms of them by four-dimensional electric-magnetic duality.
Explicitly, given the intersection form $\eta_{ij}$ on $H^3\otimes H^3$,
we have
\begin{equation} \label{eq:emdual}
dA_i = \eta_{ij} *_4~ dA_j
\end{equation}
where $*_4$ denotes the Hodge star in $d=4$.  Thus we have $h^{2,1}+1$
independent vector potentials.
One of these sits in
the $N=2$ supergravity multiplet, and the rest
are the correct number to pair with
the complex structure moduli.

We now consider $1/2$ BPS black hole
solutions of this four dimensional $N=2$ theory.  
Choosing any $S^2$ which surrounds the horizon, we can define the charge
$\gamma$ as the class in $H^3(M,\BZ)$ which reproduces the corresponding
magnetic charges
$$
Q_i = \frac{1}{2\pi}\int_{S^2} dA_i \equiv \int_M \omega_i \wedge \gamma .
$$
Using \eq{emdual}, this includes all charges.

One can show that 
the mass $M$ of any charged object in supergravity 
satisfies a \bdef{BPS bound},
\begin{equation} \label{eq:BPS}
M^2 \ge |Z(\gamma;z)|^2 .
\end{equation}
The quantity $|Z(\gamma;z)|^2$, defined in \eq{bound}, 
depends explicitly on $\gamma$, and implicitly on the
complex structure moduli $z$ through $\Omega$.
A \bdef{$1/2$ BPS solution} by definition saturates this bound.

We now explain the ``attractor paradox.''  According to
Bekenstein and Hawking, the entropy of any black hole is proportional
to the area of its event horizon.  This area can be found by finding
the black hole as an explicit solution of four-dimensional
supergravity, which clearly depends on the charge $\gamma$.
In fact, we must fix boundary conditions for all the fields at infinity,
in particular the complex structure moduli, to get a particular 
black hole solution.
Now, normally varying the boundary conditions varies all the data of a
solution in a continuous way.  On the other hand, if the entropy has
any microscopic interpretation as the logarithm of the number of
quantum states of the black hole, one would expect $e^S$ to be
integrally quantized.  Thus, it must remain fixed as the 
boundary conditions on complex
structure moduli are varied, in contradiction with naive
expectations for the area of the horizon, and seemingly contradicting
Bekenstein and Hawking.

The resolution of this paradox is the \bdef{attractor mechanism}.  
Let us work in coordinates
for which the four-dimensional metric takes the form
$$
ds^2 = -f(r) dt^2 + dr^2 + \frac{A(r)}{4\pi} d\Omega_{S^2}^2 .
$$
With some work, one can see
that in the $1/2$ BPS case, the equations of motion imply that
as $r$ decreases, the complex structure moduli $z$ follow
gradient flow with respect to $|Z(\gamma,z)|^2$ in \eq{BPS},
and the area $A(r)$ of an $S^2$ at radius $r$ decreases.
Finally, at the horizon, $z$ reaches a value $z_*$ at which $|Z(\gamma,z_*)|^2$
is a local minimum, and the area of the event horizon is 
$A=4\pi|Z(\gamma,z_*)|^2$.  Since $z_*$ is determined by minimization,
this area will not change under small variations of the initial $z$,
resolving the paradox.

A little algebra shows that the problem of finding 
non-zero critical points of $|Z(\gamma,z)|^2$, is equivalent to
that of finding
critical points $D_i Z=0$ of the period associated to $\gamma$,
\begin{equation} \label{eq:cc}
Z = \int_M \gamma \wedge \Omega
\end{equation}
usually called 
the \bdef{central charge}, with respect to the covariant derivative
\begin{equation}\label{eq:natder}
D_i Z = \partial_i Z + (\partial_i K) Z .
\end{equation}
Here
\begin{equation} \label{eq:CYK}
e^{-K} \equiv \int \Omega\wedge \bar\Omega .
\end{equation}
The mathematical significance of this rephrasing is that
$K$ is a K\"ahler potential for the Weil-Peterson K\"ahler 
metric on $\CM_c(M)$, with K\"ahler form
$\omega=\partial\bar\partial K$, and
\eq{natder} is the unique connection on $H^{(3,0)}(M,\BC)$
regarded as a line bundle over $\CM_c(M)$, whose curvature is $-\omega$.
These facts can be used to show that $D_i\Omega$ provides a basis for
$H^{(2,1)}(M,\BC)$, so that the critical point condition forces the
projection of $\gamma$ on $H^{(2,1)}$ to vanish.
This justifies our original definition \eq{att}.

\section{Flux vacua in IIb string theory}

We will not describe our second problem in
as much detail, but just give the analogous final formulation.
In this problem, a ``choice of flux'' is a pair of elements of
$H^3(M,\BZ)$, or equivalently a single element 
\begin{equation} \label{eq:fluxq}
F \in H^3(M,\BZ \oplus \tau \BZ) ,
\end{equation}
where $\tau\in\CH\equiv \{\tau\in\BC|\Im\tau>0\}$ 
is the so-called ``dilaton-axion.''

A \bdef{flux vacuum} is then a
choice of complex structure $J$ and $\tau$ for which
\begin{equation} \label{eq:fluxvac}
F \in H^{3,0}_J(M,\BC) \oplus H_J^{1,2}(M,\BC) .
\end{equation}
Now we have $h^{2,1}+h^{0,3}=h^{2,1}+1$ complex conditions on the 
joint choice of $h^{2,1}$ complex structure moduli and $\tau$, so
this condition also picks out special points, now in $\CM_c\times\CH$.

The critical point formulation of this problem is that of finding
critical points of
\begin{equation}\label{eq:defZflux}
W = \int \Omega \wedge F
\end{equation}
under the covariant derivatives \eq{natder} and
$$
D_\tau W = \partial_\tau W + (\partial_\tau W) Z
$$
with $K$ the sum of \eq{CYK} and
the K\"ahler potential $-\log\Im\tau$ 
for the metric
on the upper half plane of constant curvature $-1$.

This is a sort of complexified version of the previous problem and
arises naturally in IIb compactification by postulating a non-zero
value $F$ for a certain three-form gauge field strength, the \bdef{flux}.
The quantity \eq{defZflux} is the superpotential of the resulting
$N=1$ supergravity theory, and it is a standard fact in this context
that supersymmetric vacua (critical points of the effective potential)
are critical points of $W$ in the sense we just stated.

We can again pose the question of finding the distribution of flux
vacua in $\CM_c(M) \times \CH$.  Besides $|W|^2$, which physically 
is one of the contributions to the vacuum energy,
we can also use the ``length of the flux'' 
\begin{equation} \label{eq:defL}
L = \frac{1}{\Im\tau}\int \Re F \wedge \Im F
\end{equation}
as a control parameter, and count flux vacua for which $L\le L_{max}$.
In fact, this parameter arises naturally in the actual IIb problem, as the
``orientifold three-plane charge.''  

What makes this problem particularly interesting physically is that
it (and its analogs in other string theories) may bear on the solution
of the cosmological constant problem.  This begins with Einstein's famous
observation that the equations of general relativity admit
a one parameter generalization,
$$
R_{\mu\nu} - \half g_{\mu\nu} R = 8\pi T_{\mu\nu} + \Lambda g_{\mu\nu} .
$$
Physically, the cosmological constant $\Lambda$ is the vacuum energy,
which in our flux problem takes the form $\Lambda = \ldots - 3|W|^2$
(the other terms are inessential for us here).

Cosmological observations tell us that $\Lambda$ is very small, of the
same order as the energy of matter in the present era, about
$10^{-122} M_{Planck}^4$ in Planck units.  However, in a generic
theory of quantum gravity, including string theory, quantum effects
are expected to produce a large vacuum energy, {\it a priori} of order
$M_{Planck}^4$.  Finding an explanation for why the theory of our
universe is in this sense non-generic is the cosmological constant
problem.

One of the standard solutions of this problem is the ``anthropic
solution,'' initiated in work of Weinberg and others, and discussed in
string theory in (Bousso and Polchinski, 2000).  Suppose that we are
discussing a theory with a large number of vacuum states,
all of which are otherwise candidates to describe our universe, but
which differ in $\Lambda$.  If the number of these vacuum states were
sufficiently large, the claim that a few of these states realize a
small $\Lambda$ would not be surprising.  But one might still feel a need to
explain why our universe is a vacuum with small $\Lambda$, and not one
of the multitude with large $\Lambda$.

The anthropic argument is that, according to accepted models
for early cosmology, if the value of $|\Lambda|$ were even $100$ times
larger than what is observed, galaxies and stars could not form.
Thus, the known laws of physics guarantee that we
will observe a universe with $\Lambda$ within this bound;
it is irrelevant whether other possible vacuum states
``exist'' in any sense.  

While such anthropic arguments are controversial, one can avoid
them in this case by simply asking whether or not any vacuum state
fits the observed value of $\Lambda$.  Given a precise definition
of vacuum state, this is a question of mathematics.  Still, answering it for
any given vacuum state is extremely difficult, as it would require
computing $\Lambda$ to $10^{-122}$ precision.  But it is not out of
reach to argue that out of a large number of vacua, some of them are
expected to realize small $\Lambda$.  For example, if we could show
that the number of otherwise physically acceptable vacua was larger
than $10^{122}$, and that the distribution of $\Lambda$ among these
was approximately uniform over the range
$(-M_{Planck}^4,M_{Planck}^4)$, we would have made a good case for
this expectation.

This style of reasoning can be vastly generalized, and given favorable
assumptions about the number of vacua in a theory, could lead to
falsifiable predictions independent of any {\it a priori} assumptions
about the choice of vacuum state (Douglas, 2003).

\section{Asymptotic counting formulae}

We have just defined two classes of physically preferred points in
the complex structure moduli space of Calabi-Yau threefolds, the
attractor points and the flux vacua.  Both have simple definitions in
terms of Hodge structure, \eq{att} and \eq{fluxvac}, and both are
also critical points of integral periods of the holomorphic three-form.

This second phrasing of the problem suggests the following language.
We define a \bdef{random period} of the holomorphic three-form to be
the period for a randomly chosen cycle in $H_3(M,\BZ)$ of the types
we just discussed (real or complex, and with the appropriate control
parameters).  We are then interested in the expected distribution of
critical points for a random period.  This brings our problem into
the framework of random algebraic geometry.

Before proceeding to use this framework, let us first point out some
differences with the toy problems we discussed.  First, while \eq{cc}
and \eq{defZflux} are sums of the form \eq{fsum}, we take not an
orthonormal basis but instead a basis $s_i$ of integral periods of
$\Omega$.  Second, the coefficients $c_i$ are not normally distributed
but instead drawn from a discrete uniform distribution, {\it i.e.}
correspond to a choice of $\gamma$ in $H^3(M,\BZ)$ or $F$ as in
\eq{fluxq}, satisfying the bounds on $|Z|$ or $L$.  Finally, we do not
normalize the distribution (which is thus not a probability measure)
but instead take each choice with unit weight.

These choices can of course be modified, but are made in
order to answer the question, how many attractor points (or flux
vacua) sit within a specified region of moduli space.  
The answer
we will get is a density $\mu(Z_{max})$ or $\mu(L_{max})$
on moduli space, such that as the
control parameter becomes large, the number of critical points
within a region $R$ asymptotes to
$$
\CN(R;Z_{max}) \sim \int_R \mu(Z_{max}) .
$$

The key observation is that to get such asymptotics, we can start with
a Gaussian random element of $H^3(M,\BR)$ (or flux).
In other words, we neglect the integral quantization of the
charge or flux.  Intuitively, this might be expected to make little
difference in the limit that the charge or flux is large, and in fact
one can prove that this simplification reproduces the leading large $L$
or $|Z|$ asymptotics for the density of critical points, using standard
ideas in lattice point counting.

This justifies starting with a two-point function like \eq{twopoint}.
While the integral periods $s_i$ of $\Omega$ 
can be computed in principle (and have been in many examples) by
solving a system of linear PDE's, the Picard-Fuchs equations,
it turns out that one does not need such detailed results.  Rather,
one can use the following ansatz for the two-point function,
\begin{eqnarray*} \label{eq:Ktwop}
G(z_1,\zb_2) &= \sum_{I=1}^{b_3} \eta^{IJ} s_I(z_1) s_J^*(\zb_2) \\
 &= \int_M \Omega(z_1) \wedge \bar\Omega(\zb_2) \\
 &= \exp -K(z_1,\zb_2) .
\end{eqnarray*}
In words, the two-point function is
the formal continuation of the K\"ahler potential on $\CM_c(M)$ to
independent holomorphic and antiholomorphic variables.  This incorporates
the quadratic form appearing in \eq{defL} and can be used to count
sections with such a bound.

We can now follow the same strategy as before, by introducing an expected
density of critical points,
\begin{equation} \label{eq:expcrit}
d\mu(z) = \bigvev{\delta^{(n)}(D_i s(z))\delta^{(n)}(\Dbar_i \sb(\zb))
 \ |\det_{1\le i,j\le 2n} H_{ij}| },
\end{equation}
where the ``complex Hessian'' $H$ is the $2n\times 2n$ matrix
of second derivatives 
\begin{equation}\label{eq:comH}
H \equiv \left(\begin{matrix} 
 \partial_i \Dbar_{\jb} \sb(z)&
 \partial_i D_{j} s(z) \\
 \pb_\ib \Dbar_{\jb} \sb(z)&
 \pb_\ib D_{j} s(z) \end{matrix}\right)
\end{equation}
(note that $\partial Ds=DDs$ at a critical point).  One can then
compute this density along the same lines.  The holomorphy of $s$
implies that $\partial_i \Dbar_\jb s = \omega_{i\jb} s$, which is
one simplification.  Other geometric simplifications follow from
the fact that \eq{expcrit} depends only on $s$ and a finite number of
its derivatives at the point $z$.  

For the attractor problem, using the identity 
$$
D_i D_j s = \CF_{ijk} \omega^{k\kb} \Dbar_\kb s = 0 ,
$$
from
special geometry of Calabi-Yau threefolds,
the Hessian becomes trivial, and $\det H=|s|^{2n}$.  One thus finds 
(Denef and Douglas, 2004) that
the asymptotic density of attractor points with large $|Z|\le Z_{max}$ in
a region $R$ is
$$
\CN(R,|Z|\le Z_{max}) \sim \frac{2^{n+1}}{(n+1)\pi^n} Z_{max}^{n+1}
 \cdot {\rm vol}(R)
$$
where ${\rm vol}(R)=\int_R\omega^n/n!$ is the volume of $R$ in the
Weil-Peterson metric.  The total volume is known to be finite for Calabi-Yau
threefold moduli spaces, and thus so is the number of attractor points
under this bound.  

The flux vacuum problem is complicated by the fact that $DDs$ is non-zero
and thus the determinant of the Hessian does not take a definite sign, and
implementing the absolute value in \eq{expcrit} is nontrivial.  The
result  (Douglas, Shiffman and Zelditch, 2004) is
$$
\mu(z) \sim \frac{1}{b_3!\sqrt{\det \Lambda(z)}}
 \int_{\CH(z)\times\BC} |\det (HH^*-|x|^2\cdot{\bf 1})|
 e^{H^t\Lambda(z)^{-1}H-|x|^2} dH dx
$$
where $\CH(z)$ is the subspace of Hessian matrices \eq{comH}
obtainable from periods at the point $z$, and $\Lambda(z)$ is
a covariance matrix computable from the period data.

A simpler lower bound for the number of solutions can be obtained by 
instead computing the \bdef{index density}
\begin{equation} \label{eq:expind}
\mu_I(z) = \bigvev{\delta^{(n)}(D_i s)\delta^{(n)}(\Dbar_i \sb)
 \ \det_{1\le i,j\le 2n} H_{ij} },
\end{equation}
so-called because it weighs the vacua with a Morse-Witten sign factor.
This admits a simple explicit formula (Ashok and Douglas, 2004),
\begin{equation} \label{eq:ad}
I_{vac}(R,L\le L_{max}) \sim
  \frac{(2\pi L_{max})^{b_3}}{\pi^{n+1} b_3!} \cdot 
\int_R   \det(\CR+\omega\cdot 1) ,
\end{equation}
where $\CR$ is the $n+1\times n+1$ dimensional matrix of curvature
two-forms for the Weil-Peterson metric.  

One might have guessed at this density by the following reasoning.  If
$s$ had been a single-valued section on a compact $\CM_c$ (it is not),
topological arguments determine the total index to be
$[c_{n+1}(\CL\otimes T^*\CM)]$, and this is the simplest density
constructed solely from the metric and curvatures in the same
cohomology class.

It is not in general known whether this integral over Calabi-Yau
moduli space is finite, though this is true in examples studied so
far.  One can also control $|W|^2$ as well as other observables, and
one finds that the distribution of $|W|^2$ among flux vacua is
to a good approximation uniform.  Considering explicit examples, the
prefactor in \eq{ad} is of order $10^{100}$--$10^{300}$, so assuming
that this factor dominates the integral, we have justified the
Bousso-Polchinski solution to the cosmological constant problem in
these models.

The finite $L$ corrections to these formulae can be estimated using
van der Corput techniques, and are suppressed by better than the
naive $L^{-1/2}$ or $|Z|^{-1}$ one might have expected.  However the
asymptotic formulae for the numbers of flux vacuum break down in
certain limits of moduli space, such as the large complex structure
limit.  This is because \eq{defL} is an indefinite quadratic form,
and the fact that it bounds the number of solutions at all is 
somewhat subtle.  These points are discussed at length in
(Douglas, Shiffman and Zelditch, 2005).

Similar results have been obtained for a wide variety of flux vacuum
counting problems, with constraints on the value of the effective
potential at the minimum, on the masses of scalar fields, on scales of
supersymmetry breaking, and so on.  And in principle, this is just the
tip of an iceberg, as the study of more or less any class of superstring
vacua leads to similar questions of counting and distribution,
less well understood at present.  Some of these are discussed in 
(Douglas, 2003; Acharya et al 2005; Denef and Douglas 2005;
Blumenhagen et al 2005).

\section*{Further reading}

For background on random algebraic geometry and some of its other
applications, as well as references in the text not listed here,
consult Edelman and Kostlan, 1995 and Zelditch, 2001.
The attractor problem is discussed in Ferrara et al 1995 and
Moore, 2004, while IIb flux vacua were introduced in Giddings,
Kachru and Polchinski 2002.  Background on Calabi-Yau manifolds
can be found in Cox and Katz 1999 and Gross et al 2003.

B.~S.~Acharya, F.~Denef and R.~Valandro,
Statistics of M theory vacua,
JHEP {\bf 0506}, 056 (2005).

S. Ashok and M. R. Douglas, 
Counting Flux Vacua, JHEP 0401 (2004) 060.

R.~Blumenhagen, F.~Gmeiner, G.~Honecker, D.~Lust and T.~Weigand,
The statistics of supersymmetric D-brane models,
Nucl.\ Phys.\ B {\bf 713}, 83 (2005)

R. Bousso and J.  Polchinski,
Quantization of four-form fluxes and dynamical neutralization of
the cosmological constant,  
JHEP 06, 06 (2000).

D. A. Cox and S. Katz,
{\it Mirror symmetry and algebraic geometry},
American Mathematical Society, Providence, RI, 1999.

F.~Denef and M.~R.~Douglas,
Distributions of flux vacua,
JHEP {\bf 0405}, 072 (2004).

F.~Denef and M.~R.~Douglas,
Distributions of nonsupersymmetric flux vacua,
JHEP {\bf 0503}, 061 (2005).

M. R. Douglas,  
The statistics of string/M theory vacua,
JHEP 5, 046 (2003).

M. R. Douglas, B. Shiffman and S. Zelditch,
Critical Points and supersymmetric vacua I,  
Comm. Math. Phys. 252 (2004), no. 1-3, 325--358;
II: Asymptotics, and III: String/M Models, to appear.

A. Edelman and E. Kostlan,
How many zeros of a random polynomial are real?
Bull. Amer. Math. Soc. (N.S.) 32 (1995) 1-37.

S.~Ferrara, R.~Kallosh and A.~Strominger,
$N=2$ extremal black holes,
Phys.\ Rev.\ D {\bf 52}, 5412 (1995).

S.B. Giddings, S.  Kachru, and J. Polchinski,
Hierarchies from fluxes in string compactifications,
Phys. Rev. D (3) 66 (2002), no. 10, 106006.

M. Gross, D. Huybrechts, and D. Joyce, 
{\it Calabi-Yau Manifolds and Related Geometries}, 
Springer Universitext, Springer, New York, 2003.

G.~W.~Moore,
Les Houches lectures on strings and arithmetic,
arXiv:hep-th/0401049.

S. Zelditch, 
From random polynomials to symplectic geometry,
in {\it XIIIth International Congress of Mathematical Physics,}
International Press (2001), 367-376.

S. Zelditch, Random complex geometry and vacua, or: How to count
universes in string/M theory, 2005 preprint.

\end{document}